

\input phyzzx

\let\refmark=\NPrefmark 
\def\define#1#2\par{\def#1{\Ref#1{#2}\edef#1{\noexpand\refmark{#1}}}}
\def\con#1#2\noc{\let\?=\Ref\let\<=\refmark\let\Ref=\REFS
         \let\refmark=\undefined#1\let\Ref=\REFSCON#2
         \let\Ref=\?\let\refmark=\<\refsend}

\define\RNARAIN
K. Narain, Phys. Lett. {\bf B169} (1986) 41.

\define\RNSW
K. Narain, H. Sarmadi and E. Witten, Nucl. Phys. {\bf B279} (1987) 369.

\define\ROLIVE
C. Montonen and D. Olive, Phys. Lett. {\bf B72} (1977) 117;
H. Osborn, Phys. Lett. {\bf B83} (1979) 321.

\define\RIBANEZ
A. Font, L. Ibanez, D. Lust and F. Quevedo, Phys. Lett. {\bf B249} (1990)
35;
J. Harvey and J. Liu, Phys. Lett. {\bf B268} (1991) 40;
S.J. Rey, Phys. Rev.{\bf D43} (1991) 526.

\define\RMONOPOLE
X. G. Wen and E. Witten, Nucl. Phys. {\bf B261} (1985) 651;
R. Rohm and E. Witten, Ann. Phys. (NY) {\bf 170} (1986) 454;
T. Banks, M. Dine, H. Dijkstra and W. Fischler, Phys. Lett. {\bf B212}
(1988) 45;
J. Harvey and J. Liu, in ref.\RIBANEZ;
R. Khuri, preprints CTP/TAMU-33/92 (hep-th/9205051), CTP/TAMU-35/92
(hep-th/9205081);
J. Gaunlett, J. Harvey and J. Liu, preprint EFI-92-67 (hep-th/9211056).

\define\RSTROM
A. Strominger, Nucl. Phys. {\bf B343} (1990) 167; C. Callan, J. Harvey and
A. Strominger, Nucl. Phys. {\bf B359} (1991) 611; {\bf B367} (1991) 60;
preprint EFI-91-66.

\define\RDUFF
M. Duff, Class. Quantum Grav. {\bf 5} (1988) 189;
M. Duff and J. Lu, Nucl. Phys. {\bf B354} (1991) 129, 141; {\bf B357}
(1991) 354;
Phys. Rev. Lett. {\bf 66} (1991) 1402;
Class. Quantum Grav. {\bf 9} (1991) 1;
M. Duff, R. Khuri and J. Lu, Nucl. Phys. {\bf B377} (1992) 281.
J. Dixon, M. Duff and J. Plefka, preprint CTP-TAMU-60/92
(hepth@xxx/9208055).

\define\RTOL
I. Pesando and A. Tollsten, Phys. Lett. {\bf B274} (1992) 374.

\define\RGAZU
M. Gaillard and B. Zumino, Nucl. Phys. {\bf B193} (1981) 221.

\define\RODD
S. Ferrara, J. Scherk and B. Zumino, Nucl. Phys. {\bf B121} (1977) 393;
E. Cremmer, J. Scherk and S. Ferrara, Phys. Lett. {\bf B68} (1977) 234;
{\bf B74} (1978) 61;
E. Cremmer and J. Scherk, Nucl. Phys. {\bf B127} (1977) 259;
E. Cremmer and B. Julia, Nucl. Phys.{\bf B159} (1979) 141;
M. De Roo,  Phys. Lett. {\bf B156}
(1985) 331;
E. Bergshoef, I.G. Koh and E. Sezgin, Phys. Lett. {\bf B155} (1985) 71;
M. De Roo and P. Wagemans, Nucl. Phys. {\bf B262} (1985) 646;
L. Castellani, A. Ceresole, S. Ferrara, R. D'Auria, P. Fre and E. Maina,
Nucl. Phys. {\bf B268} (1986) 317; Phys. Lett. {\bf B161} (1985) 91;
S. Cecotti, S. Ferrara and L. Girardello, Nucl. Phys. {\bf B308} (1988)
436;
M. Duff, Nucl. Phys. {\bf B335} (1990) 610.

\define\RDEROO
M. De Roo, Nucl. Phys. {\bf B255} (1985) 515.

\define\RSTW
A. Shapere, S. Trivedi and F. Wilczek, Mod. Phys. Lett. {\bf A6}
(1991) 2677.

\define\RDUALITY
A. Sen, preprint TIFR-TH-92-41 (hepth@xxx/9207053).

\define\RDYON
A. Sen, preprint TIFR-TH-92-46 (hepth@xxx/9209016) (to appear in Phys.
Lett. B).

\define\RSCHWARZ
J. Schwarz, preprint CALT-68-1815 (hepth/9209125).

\define\RORTIN
T. Ortin, preprint SU-ITP-92-24 (hepth@xxx/9208078).

\define\RDGHR
A. Dabholkar, G. Gibbons, J. Harvey and F.R. Ruiz, Nucl. Phys. {\bf B340}
(1990) 33;
A. Dabholkar and J. Harvey, Phys. Rev. Lett. {\bf 63} (1989) 719.

\define\RSTRING
A. Sen, preprint TIFR-TH-92-39 (hep-th/9206016) (to appear in Nucl.
Phys. B)

\define\RCOMP
S. Ferrara, C. Kounnas and M. Porrati, Phys. Lett. {\bf 181B} (1986) 263;
M.V. Terentev, Sov. J. Nucl. Phys. {\bf 49} (1989) 713.

\define\RGREENE
B. Greene, A. Shapere, C. Vafa and S. Yau, Nucl. Phys. {\bf B337} (1990)
1.

\define\RSIKIVIE
P. Sikivie, Phys. Lett. {\bf 137B} (1984) 353.

\def\vl{\vec l}
\def\a{\vec\alpha}
\def\b{\vec\beta}
\def\Z{Z}
\def\R{R}
\def\bz{\bar z}
\def\p{\partial}
\def\ta{\tilde a}
\def\tb{\tilde b}
\def\tc{\tilde c}
\def\td{\tilde d}
\def\tg{\tilde g}
\def\vqe{\vec Q_e}
\def\vqm{\vec Q_m}
\def\mo{M^{(0)}}
\def\loo{\lambda_1^{(0)}}
\def\lo{\lambda^{(0)}}
\def\lpo{\lambda^{\prime(0)}}
\def\lto{\lambda_2^{(0)}}
\def\lpoo{\lambda_1^{\prime(0)}}
\def\lpto{\lambda_2^{\prime(0)}}
\def\vF{\vec F}
\def\vtF{\vec{\tilde F}}
\def\vE{\vec E}
\def\vB{\vec B}
\def\vq{\vec q}

{}~\hfill\vbox{\hbox{TIFR-TH-93-03}\hbox{hep-th/9302038}\hbox{February,
1993}}\break

\title{SL(2, Z) DUALITY AND MAGNETICALLY CHARGED STRINGS}

\author{Ashoke Sen\foot{e-mail address: SEN@TIFRVAX.BITNET}}

\address{Tata Institute of Fundamental Research, Homi Bhabha Road, Bombay
400005, India}

\abstract

Postulate of SL(2,\Z) invariance of toroidally compactified heterotic
string theory in four dimensions implies the existence of new string (dual
string) states carrying both, electric and magnetic charges.
In this paper we study the interaction between these dual strings.
In particular, we consider scattering of two such strings in the limit
where one string passes through the other string without touching it, and
show that such a scattering leads to the exchange of a fixed amount of
electric and magnetic charges between the two strings.

\endpage

\chapter{Introduction}

Following earlier
ideas\con\ROLIVE\RIBANEZ\RSTROM\RDUFF\RTOL\RGAZU\RODD\RDEROO\RSTW\noc\ we
have
proposed recently\con\RDUALITY\RDYON\noc\ that the toroidally compactified
heterotic string theory\RNARAIN\RNSW\ in four dimensions may be invariant
under an
SL(2,\Z) group of transformations.
These transformations mix electric and magnetic fields, and at the same
time act non-trivially on the axion-dilaton field, thereby interchanging
the strong and weak coupling limits of the theory.
Further work in this direction was reported in ref.\RSCHWARZ.
SL(2,\R) symmetry was used in refs.\RSTW\RDUALITY\RORTIN\ to generate
magnetically
charged black hole solutions in string theory.

In order that the full string theory has SL(2,\Z) invariance, the theory
must contain magnetically charged states.
The allowed spectrum of electric and magnetic charges in the theory was
computed in ref.\RDYON.
A natural question to ask would be, `where do these magnetically charged
states come from?'
A partial answer to this question was provided in ref.\RDUALITY.
Following ref.\RDGHR, if we regard fundamental strings as solitons of the
effective field theory (a description which is likely to hold for states
representing long strings) then the dual strings may be constructed simply
from the SL(2,\Z) transform of these solitons.

In this paper we shall study the interaction between these dual strings,
which also includes interaction between a dual string and an ordinary
string.
In particular, we show that the force between an infinitely long straight
dual string and an
ordinary test string parallel to the dual string vanishes.
We then study the scattering of two closed dual strings when one of them
passes
through the other without touching it, and show that the result is an
exchange of a fixed amount of electric and magnetic charge between the two
strings, determined by the quantum numbers of the original string.

The plan of the paper is as follows.
In sect.2 we give a brief review of SL(2, \Z) invariance in toroidally
compactified heterotic string theory, and also discuss the relationship
between classical solutions in this theory and fundamental strings.
In sect.3 we construct the magnetically charged dual string solutions by
taking SL(2,\Z) transformation of the fundamental string solution.
In sect.4 we calculate the force between a dual string and an ordinary
test string parallel to it, and show that it vanishes.
In sect.5 we study the result of adiabatically transporting a particle,
carrying both, electric and magnetic charges, around a dual string and
show that both these charges change as a result of this transport.
In sect.6 we derive the same results by regarding the string as the
boundary of a domain wall, and calculating the electric and magnetic
charges exchanged between the particle and the domain wall as the particle
passes through the wall.
In sect.7 we use the results of sect.5 and 6 to study the scattering of
two strings.
We summarize our results in sect.8.

\chapter{Review}

We begin by giving a brief review of the duality invariance of the
effective field theory and soliton solutions in this theory representing
fundamental strings.
We consider heterotic string theory with six of its ten dimensions
compactified on a torus with constant background gauge and anti-symmetric
tensor fields.
For a generic compactification, the only massless bosonic fields in the
theory are the metric $G_{\mu\nu}$, a complex scalar field $\lambda$
representing the axion-dilaton system, a set of 28 gauge fields
$A_\mu^{(\alpha)}$ ($1\le\alpha\le 28$) which we shall denote as a 28
dimensional vector $\vec A_\mu$, and a $28\times 28$ matrix valued field
$M$ satisfying,
$$
M^T=M, ~~~~ M^T L M = L
\eqn\etwo
$$
where
$$
L=\pmatrix{ 0 & I_6 & 0\cr I_6 & 0 & 0\cr 0 & 0 & -I_{16}}
\eqn\ethree
$$
$I_n$ being the $n\times n$ identity matrix.
In terms of these fields, the action is given by,
$$\eqalign{
S =&{1\over 32\pi}\int d^4 x\sqrt{-\det G}[ R -{1\over 2(\lambda_2)^2}
G^{\mu\nu}\p_\mu\lambda\p_\nu\bar\lambda -\lambda_2\vec F^T_{\mu\nu}.LML.
\vec F^{\mu\nu} \cr
& +\lambda_1\vec F^T_{\mu\nu}.L.\vec{\tilde F}^{\mu\nu} +{1\over 8}
G^{\mu\nu}
Tr(\p_\mu M L\p_\nu M L)]\cr
}
\eqn\eone
$$
where
$$
\vec F_{\mu\nu} =\p_\mu \vec A_\nu - \p_\nu\vec
A_\mu
\eqn\efour
$$
The equations of motion derived from this action are invariant under the
SL(2,\R) transformation\RDUALITY:
$$\eqalign{
& \lambda\to {a\lambda + b\over c\lambda + d}, ~~~~ \vec F_{\mu\nu}\to
c\lambda_2 ML . \vec{\tilde F}_{\mu\nu} + (c \lambda_1 + d) \vec
F_{\mu\nu}\cr
& M\to M, ~~~~ G_{\mu\nu}\to G_{\mu\nu}\cr
}
\eqn\efive
$$
where $a,b,c,d$ are real numbers.
However, quantum corrections due to instantons break the SL(2,\R)
invariance to at most
SL(2,\Z) invariance, for which,
$$
a,b,c,d\in \Z, ~~~~ ad-bc=1
\eqn\esix
$$

The equations of motion derived from the action \eone\ has a string like
classical solution\RDGHR, given by,
$$\eqalign{
\lambda =& {1\over 2\pi i} \ln {z\over r_0}\cr
ds^2 =& -dt^2 +(dx^3)^2 -{1\over 2\pi} \ln{r\over r_0} dz d\bz\cr
}
\eqn\eseven
$$
This describes a fundamental string lying along the $x^3$ direction.
$z=x^1 +i x^2$ denotes the complex coordinate transverse to the string.
The core of the string in this case is located at $z=0$.
This solution can be shown to be invariant under eight of the sixteen
global supersymmetry generators\RDGHR.

Note that the solution is sensible only in the region $r<r_0$, where $r_0$
is an arbitrary length scale.
{}From physical consideration we see that
$r_0$ should be taken to be of the order of the overall size of the closed
string loop.
Only for $r<<r_0$ the string looks like a straight string, and eq.\eseven\
gives a good description of the field configuration in this region.
$r_0$ also contains information about the asymptotic value of $\lambda_2$;
for closed string loops of the same size, $r_0$ takes different values for
different asymptotic values of the field $\lambda_2$.

The solution has several zero modes. First of all there are two bosonic
zero modes which simply correspond to shifting the location of the
core of the string in the $x^1-x^2$ plane. There are eight fermionic
zero modes which correspond to supersymmetry transformation of the
solution with the eight broken global supersymmetry generators. These
supersymmetry generators are chiral with respect to the gamma matrices
associated with the $t-x^3$ direction\RDGHR\RSTRING, hence the
corresponding fermionic
zero modes are also chiral. In particular for the solution given in
eq.\eseven, they turn out to be right chiral. Finally, there are 28 bosonic
zero modes generated by $O(7,23)$ deformation of the solution as discussed
in refs.\RDUALITY\RSTRING. The parameters labelling the deformed solution
may be identified as the charge per unit length carried by the string
corresponding to the 28 gauge fields. To first order in the deformation
parameters $q^{(I)}$ ($13\le I\le 28$), $p^{(m)}$, $l^{(m)}$ ($1\le m\le
6$), the
solution deformed by these charge zero modes is given by,
$$\eqalign{
\lambda =& {1\over 2\pi i}\ln{z\over r_0}\cr
ds^2 =& -dt^2 + (dx^3)^2 -{1\over 2\pi} \ln{r\over r_0} dz d\bar z\cr
F^{(I)}_{-zt}=& F^{(I)}_{-z3} = {q^{(I)}\over z} {1\over
(\ln(r/r_0))^2} ~~{\rm for}~13\le I\le 28\cr
F^{(m)}_{-zt}-F^{(m+6)}_{-zt} =& F^{(m)}_{-z3}-F^{(m+6)}_{-z3} ={\sqrt 2
p^{(m)}\over z} {1\over (\ln(r/r_0))^2}~~{\rm for}~ 1\le m\le 6\cr
F^{(m)}_{-zt}+F^{(m+6)}_{-zt} =& -(F^{(m)}_{-z3}+F^{(m+6)}_{-z3}) ={\sqrt
2 l^{(m)}\over z} {1\over (\ln(r/r_0))^2}~~{\rm for}~ 1\le m\le 6\cr
F^{(\alpha)}_{-\bar z t}=& F^{(\alpha)}_{-\bar z 3} =0~~{\rm for~} 1\le
\alpha\le 28\cr
M =& I\cr
}
\eqn\eeight
$$
which generalizes the solution given in ref.\RSTRING\ where only the
parameters $q^{(I)}$ were present.
Here
$$
\vec F_{\pm \mu \nu} = -ML . \vec F_{\mu\nu}\pm i\vec{\tilde F}_{\mu\nu}
\eqn\eeightaa
$$
$q^{(I)}$, $p^{(m)}$ measure the charge per unit length, as well as the
current in the $- x^3$ direction associated with the gauge fields
$A_\mu^{(I)}$
($13\le I\le 28$) and $(A_\mu^{(m)} - A_\mu^{(m+6)})/\sqrt 2$ ($1\le m\le
6$)
respectively, and $l^{(m)}$ measure the charge per unit length, and the
current in the $x^3$ direction associated with the gauge field $(A_\mu^m
+ A_\mu^{(m+6)})/\sqrt 2$ ($1\le m\le 6$).

The collective excitation of
the string may be described by making the parameters labelling these
deformations functions of $t$ and $x^3$.
In particular, when we make $q^{(I)}$, $p^{(m)}$ and $l^{(m)}$ functions
of $t$ and
$x^3$, charge conservation implies that,
$$
(\p_t - \p_3) q^{(I)} = (\p_t-\p_3) p^{(m)} = (\p_t + \p_3) l^{(m)} =0
\eqn\enine
$$
therby showing that $q^{(I)}$ and $p^{(m)}$ denote left moving
coordinates and $l^{(m)}$ denote right moving coordinates.

The set of all the collective excitations of the string can easily be seen
to be in one to one correspondence with the dynamical degrees of freedom
of the fundamental string in the static gauge.
This leads to the hypothesis that the quantization of these collective
coordinates will reproduce the full spectrum of states in the string
theory.
Although formally this may be an exact result after taking into account
the correction to the action due to the higher derivative terms, in
practice the usefulness of this hypothesis is limited to states associated
with long strings.

\chapter{Dual Strings}

In ref.\RDYON\ we have indicated that the allowed spectrum of
electric and magnetic charges in string theory is consistent with the
SL(2,\Z) invariance of the theory. This, however, does not answer the
question as to where the magnetically charged states, that are necessary
for SL(2,\Z) invariance of the spectrum, come from. In this section we
shall try to partially answer this question.

The answer in fact lies in the hypothesis stated at the end of the last
section.
Since according to this hypothesis, string states can be regarded as
collective excitations of a classical solution in the effective field
theory, the magnetically charged string states must come from the
collective excitations of the SL(2,\Z)
transform of this classical solution. To make this more concrete, let us
first write down the SL(2,\Z) transform of the solution given in
eq.\eseven\ by the element $g=\pmatrix{a & b\cr c & d\cr}$:
$$\eqalign{
\lambda =&{ a\ln (z/r_0) + 2\pi i b\over c\ln (z/r_0) + 2\pi i d}\cr
ds^2 =& - dt^2 + (dx^3)^2 -{1\over 2\pi}\ln(r/r_0) dz d\bz\cr
}
\eqn\eten
$$
Note that as we go around the string, $\lambda$ changes to
$$
(\ta \lambda + \tb)/ (\tc\lambda +\td)
\eqn\eeleven
$$
where,
$$
\tg\equiv \pmatrix{\ta & \tb\cr \tc & \td\cr} = \pmatrix{ a & b\cr c &
d\cr} T \pmatrix{a & b\cr c & d\cr}^{-1}
=\pmatrix{1 - ac & a^2\cr -c^2 & 1+ac\cr}
\eqn\etwelve
$$
where
$$
T = \pmatrix{1 & 1\cr 0 & 1\cr}
\eqn\etwelvea
$$
The zero modes of this solution can be constructed in the following way.
Instead of trying to write down the deformed solution directly, we can
simply take the zero mode deformation of the solution \eseven, and take
the SL(2,\Z) transform of it.
SL(2,\Z) invariance of the equations of motion will imply that the
transformed configuration is also a solution of the equations of motion,
and hence denotes the zero mode deformation of the solution given in
eq.\eten.
Quantization of these zero modes should, in turn, produce the magnetically
charged strings required for duality invariance of the theory, at least
those corresponding to long strings.
We shall not explicitly display the deformed solution here.

Note that the electrically and magnetically charged string states in a
theory with a given asymptotic value of $\lambda$ are not related to each
other by SL(2,\Z) transformation.
Instead, the magnetically charged particles in this theory are related
by SL(2, \Z) transformation to
the purely electrically charged particles in a theory with a different
asymptotic value of $\lambda$.

There is one question that must be addressed before we conclude this
section.
So far we have discussed SL(2,\Z) invariance of the theory in the case
where the fermionic background fields have been set to zero.
But the true SL(2,\Z) invariance of the theory requires SL(2, \Z)
invariance of the equations of motion even in the presence of fermionic
background fields.
In particular, this is necessary if we want to construct the fermionic
zero modes of the SL(2,\Z) transformed solution.
We shall now give an indirect proof of the SL(2, \Z) invariance of the
equations of motion after the inclusion of the fermionic fields.
This is done by comparing the dimensionally reduced heterotic string
theory to the $N=4$ Poincare supergravity theory coupled to abelian gauge
field multiplets as discussed in ref.\RDEROO.
It can be shown that the bosonic part of the action
given in eqs.(4.18), (4.26) of ref.\RDEROO\ is identical to the action
given in eq.\eone\ after
we make the following identification of fields\foot{
Partial results to this effect have been obtained previously in
refs.\RCOMP.}
$$
M = U OO^T U^{-1}, ~~~~~~ {i\over\lambda} = {\phi_1 -\phi_2\over \phi_1 +
\phi_2}
\eqn\esixtynine
$$
and a redefinition of the gauge fields $\vF\to U\vF$.
Here $U$ is a matrix that diagonalizes $L$:
$$
U^{-1} L U =\pmatrix{ I_6 & &\cr & -I_6 &\cr & & -I_{16}\cr}
\eqn\eseventy
$$
In eq.\esixtynine, the right hand sides of the equations contain variables
appearing in ref.\RDEROO, whereas the left hand sides of the equations
contain variables appearing in eq.\eone.
Since the bosonic part of the two actions are identical,
we have a strong evidence that the two theories are indeed
the same.
We shall proceed with the assumption that this is the case.
This assumption is further supported by the fact that both theories have
local $N=4$ supersymmetry.

In ref.\RDEROO\ it was shown that the gauge field equations of motion are
invariant under SL(2,\R) transformation even after including the fermionic
fields in this theory. This result, combined with an earlier result of
Gaillard and Zumino\RGAZU\ shows that all the field equations must be
invariant under the SL(2,\R) transformation.
This establishes the SL(2,\R) invariance of the full
set of equations of motion of the dimensionally reduced heterotic string
theory.
Similar argument has been advanced previously by Schwarz\RSCHWARZ.

\chapter{Force Exerted by a Dual String on an Ordinary String}

We now begin our study of the interaction between dual strings, and also
between a dual string and an ordinary string.
The first quantity we would like to compute is the force exerted by an
infinitely long straight dual
string on an ordinary test string kept parallel to itself some distance
away.
A similar computation for two ordinary strings parallel to each other had
yielded the answer that the net force between such strings
vanish\RDGHR\RSTRING.

We begin by writing down the action of a test string in the presence of a
background axion-dilaton-gravitational field:
$$
S_{string} = \int d^2\xi (\sqrt{-\gamma} G_{S\mu\nu}(X)\gamma^{\alpha\beta}
\p_\alpha X^\mu \p_\beta X^\nu +\epsilon^{\alpha\beta} B_{\mu\nu}(X)
\p_\alpha X^\mu \p_\beta X^\nu ) +\ldots
\eqn\efourteen
$$
where $\xi^\alpha$ and $\gamma_{\alpha\beta}$ denote the coordinates and
metric on the string world-sheet respectively, $X^\mu$ denote the
coordinates of the string, $\Phi$ denotes the dilaton field, and
$G_{S\mu\nu}$ denotes the string metric.
$\ldots$ denotes terms involving background gauge fields and world sheet
fermionic fields, which
we are setting to zero for the present analysis.
The relation between the fields appearing here and those in eq.\eone\ are
given by,
$$\eqalign{
G_{S\mu\nu} =& e^{\Phi} G_{\mu\nu}, ~~~~~~ e^{-\Phi} = \lambda_2\cr
G^{\sigma\sigma'}\p_{\sigma'}\lambda_1 =& {1\over 2} (\sqrt{-\det G})^{-1}
e^{-2\Phi}\epsilon^{\mu\nu\rho\sigma} (\p_\mu B_{\nu\rho} + {1\over 2}
\vec A_\mu^T . L . \vec
F_{\nu\rho}) \cr
}
\eqn\efifteen
$$
Let us  now consider the test string lying along the 3 direction,
$$
X^0 =\xi^0, ~~~~ X^3 =\xi^1
\eqn\esixteen
$$
with the gauge choice,
$$
\sqrt{-\gamma} \gamma^{\alpha\beta} =\eta^{\alpha\beta}
\eqn\esixteena
$$
If $Z=X^1+i X^2$ denotes the complex coordinate transverse to the string,
then the equation of motion of $Z$ that follows from the action
\efourteen\ in the background \eten\ is given by (with the help of
eqs.\efifteen)
$$
D^\alpha D_\alpha Z +\Gamma^Z_{\mu\nu}\p_\alpha X^\mu \p^\alpha X^\nu -i
{\p_{\bar Z}\lambda \over \lambda_2} G^{Z\bar Z}=0
\eqn\eseventeen
$$
The second term, which in this gauge is given by
$\Gamma^Z_{33}-\Gamma^Z_{tt}$ vanishes, as can easily be seen by computing
the
Christoffel symbol from the metric given in eq.\eten.
The last term also vanishes, since $\lambda$ given in eq.\eten\ is an
analytic function of $z$.
The net result is that the equation of motion of the coordinate $Z$ looks
like,
$$
D^\alpha D_\alpha Z =0
\eqn\eeighteen
$$
showing that there is no net force exerted on the test string.

The same analysis can also be repeated for the case where the dual string
carries electric and magnetic charge density (these are the solutions
deformed by the charge zero modes). As in the case of ref.\RSTRING, in
this case the net magnetic and electric forces exerted on the test string
cancel each other.

\chapter{Adiabatic Transport of a Charged Particle Around a Dual
String}

We shall now consider the effect of adiabatically transporting a point
particle carrying magnetic charge $\vec Q_m$ and electric charge $\vec
Q_e$ around a dual string.\foot{Similar phenomena associated with the
usual $R\to 1/R$ duality transformation was discussed in ref.\RGREENE.}
Although the results of this and the next section may be derived by
starting from the known results about the interaction of a dyon with an
axionic domain wall\RSIKIVIE\ and then making a duality transformation of
the results, we shall carry out the analysis explicitly, since this gives
a physical understanding of the interaction mechanism.
As was shown in ref.\RDYON\ the allowed spectrum of $(\vec Q_m, \vec Q_e)$
is
$$
(\vqm,\vqe)=(\mo L\b, ~(\a +\loo \b)/\lto)
\eqn\enineteen
$$
where $\loo$, $\lto$ and $\mo$ are the asymptotic values of the fields
$\lambda_1$, $\lambda_2$ and $M$ respectively, and $\a$, $\b$ are 28
dimensional vectors
belonging to an even, self dual lattice $P$ with the metric $L$.
Let us now consider a dual string that is a transform of the ordinary
string by the group element
$g=\pmatrix{a & b\cr c & d}$ and let $\tg=\pmatrix{\ta & \tb\cr \tc &
\td}$ be the corresponding element defined in eq.\etwelve.
As we adiabatically transport the particle around the string, we expect
$\a$, $\b$ to remain fixed since they take discrete values.
After we go around the string once, the background value of the field
$\lambda$ changes to
$$
\lambda' = {\ta \lambda +\tb\over \tc\lambda +\td}
\eqn\etwentyone
$$
which, in turn, implies that the electric and magnetic charge vectors of
the particle change to
$$
(\vec Q_m', \vec Q_e') =(\mo L\b, {1\over \lpto}(\a +\lpoo\b))
\eqn\etwentytwo
$$
However, the background seen by the particle is now different from the one
that was seen by it before.
As a result, we should not directly compare $(\vec Q_m', \vec Q_e')$ with
$(\vec Q_m, \vec Q_e)$.
Instead, we need to make a duality transformation of the fields and the
charges by the element,
$$
\tg^{-1}=\pmatrix{\td & -\tb\cr -\tc & \ta}
\eqn\etwentythree
$$
to state the results in the original coordinate system.
This transformation sends $\lambda'$ back to $\lambda$, and $(\a,\b)$ to
$(\a', \b')$ given by\RDYON\
$$
\pmatrix{\a'\cr -\b'} = \pmatrix{\td & -\tb\cr -\tc & \ta\cr} \pmatrix{\a
\cr -\b}
\eqn\etwentyfour
$$
Thus the final electric and magnetic charges of the particle are given by:
$$
(\vec Q_m'', \vec Q_e'') = (\mo L\b', {1\over\lto}(\a' +\loo\b'))
\eqn\etwentyfive
$$

The conservation of electric and magnetic charge implies that
the charge lost by the particle must be deposited on the string, but the
above analysis does not show explicitly how it happens.
Also, for a realistic string, once the SL(2,\Z) symmetry is broken by the
instanton corrections, the field around a string is not given by eq.\eten,
but remains equal to its vacuum value $\lo$ in most of the region of
space, and changes quickly to $\lpo=(\ta\lo+\tb)/(\tc\lo+\td)$ across a
thin domain wall bounded by the string.
We shall now derive eq.\etwentyfive\ using this more realistic picture of
strings, which will also clarify the charge exchange mechanism between the
string and the domain wall.

\chapter{Dynamics of Domain Wall Penetration}

\centerline{\singlespace
\vbox{\hbox{A}
\hbox{$\big |$\hskip -.1in $\cdot$}\hbox{$\big |$\hskip -.2in
$\cdot$}\hbox{$\big |$\hskip -.3in $\cdot$}\hbox{$\big |$\hskip -.35in
$\cdot$}\hbox{$\big |$\hskip -.4in
$\cdot$\hskip -.3in D\hskip .7in C}
\hbox{$\big |$\hskip -.35in $\cdot$}\hbox{$\big |$\hskip -.3in $\cdot$}
\hbox{$\big |$\hskip -.2in $\cdot$}\hbox{$\big |$\hskip -.1in $\cdot$}
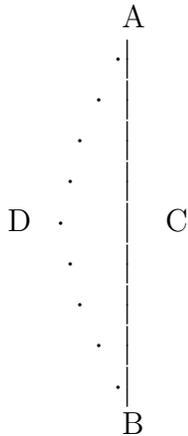
\hbox{B}\hbox{~}\hbox{~}\hbox{~\hskip -1.4in Fig.~1. Picture of a string}
}}

Let us consider the picture of the string depicted in Fig.1. The points
$A$ and $B$ in this figure denote the points at which the string intersects
the plane of the paper.
The line $C$ connecting the two points represents the intersection of the
domain wall (bounded by the string) with the plane of the paper.
On the right side of the wall $C$ the field $\lambda$ takes the value
$\lo$, whereas on the left side of the wall the field $\lambda$ takes
the value $\lpo$.
Finally, the line $D$ represents the intersection of another fictitious
wall bounded by the string with the plane of the paper.
This fictitious
wall is characterised by the choice of two different coordinate systems on
the two sides of the wall, related by the SL(2,\Z) transformation with the
group element $\tg$ such that on the left hand side of the wall $D$ the
field $\lambda$ takes value $\lo$.
In other words, the field $\lambda$ takes value $\lo$ on the right side of
$C$ and the left side of $D$, but takes the value $\lpo$ in the region
bounded by $C$ and $D$.
Note that $C$ represents a real domain wall with finite
thickness, and $\lambda$ changes continuously from $\lo$ to $\lpo$ as we
cross the wall, whereas $D$ denotes an infinitely thin fictitious wall,
and arises only because we choose to use two different coordinate systems
on the two sides of the wall.
As we shall see, the results of passage of a particle through these two
walls are completely different.
In both cases, however, as the particle crosses the wall, it deposits
certain amount of electric and magnetic charge on the wall, which
ultimately flows back to the string.

In studying the passage of charged particles through these walls, we shall
use the method used by Sikivie\RSIKIVIE\ for studying the passage of
charged particles through an axionic domain wall.
First let us consider the passage of the particle through $C$.
Both, inside, and outside the wall, the gauge fields satisfy the equation
of motion:
$$
D_\mu (\lambda_2ML\vF^{\mu\nu} -\lambda_1\vtF^{\mu\nu})=0
\eqn\etwentysix
$$
Let us now ignore all time derivatives (assuming that the motion of the
particle is slow) and define,
$$
\vF^{i0}=\vE^i,~~~~\vtF^{i0}=\vB^i
\eqn\etwentyseven
$$
We shall also assume that the energy per unit area of the wall
is small compared to $M_{Pl}^3$,
so that we can ignore the gravitational field produced by the wall.
At the same time, the thickness of the wall is taken to be small compared
to the overall size of the string, so that the variation of the
gravitational field due to the string across the wall is small.
The equation of motion \etwentysix\ and the Bianchi identity now takes the
form:
$$\eqalign{
D_i (\lambda_2 ML\vE^i-\lambda_1\vB^i)=& 0, ~~~~ \epsilon^{ijk}D_j
(\lambda_2 ML \vec B_k +\lambda_1 \vec E_k)=0\cr
D_i \vB^i=& 0, ~~~~ \epsilon^{ijk} D_j\vec E_k=0\cr
}
\eqn\etwentyeight
$$
Let us now denote by $\vB_{\perp}$ ($\vE_\perp$) and $\vB_\perp'$
($\vE_\perp'$) the components of the
magnetic (electric) fields perpendicular to the domain wall on the right
and the left side of $C$ respectively.
Similarly, we denote by $\vB_{||}$ ($\vE_{||}$) and $\vB_{||}'$
($\vE_{||}'$) the components of magnetic (electric) fields parallel to the
wall on the two sides of the wall.
In the thin wall approximation, the variation of various fields in
directions parallel to the wall are small compared to that in directions
perpendicular to the wall.
Eq.\etwentyeight\ then gives
$$\eqalign{
\vB'_\perp =& \vB_\perp, ~~~~ \lpto\mo L\vE'_\perp -\lpoo\vB'_\perp =
\lto\mo L\vE_\perp -\loo\vB_\perp \cr
\vE_{||}' =& \vE_{||}, ~~~~ \lpto\mo L \vB'_{||} +\lpoo\vE'_{||} =
\lto\mo L \vB_{||} +\loo\vE_{||}\cr
}
\eqn\ethirty
$$
Thus the total induced magnetic charge on the wall is given by,
$$
\Delta\vec Q_m ={1\over 4\pi}\int (\vB'_\perp -\vB_\perp) d^2 S =0
\eqn\ethirtyone
$$
On the other hand, the total induced electric charge is given by,
$$
\Delta\vec Q_e ={1\over 4\pi}\int (\vE'_\perp -\vE_\perp)d^2S
\eqn\ethirtytwo
$$
and is non-zero in general.

We now consider a particle with charge $(\vec Q_m, \vec Q_e)$ approaching
the wall $C$ from the right.
The electromagnetic fields due to the particle in the absence of the
domain wall $C$ are given by
$\vE^i =\vec Q_e r^i/r^3$, $\vB^i =\vec Q_m r^i/r^3$.
In order to calculate the total induced charge on the wall, we need to
calculate the electric and magnetic fields that are obtained by solving
eqs.\ethirty\ and then use eqs.\ethirtyone, \ethirtytwo.
This is done using the method of images.
Let $P$ denote the position of the incoming particle at a given instant of
time and $Q$ be its image point.
We assume that the field to the right side of $C$ is reproduced by the
original particle at the point $P$ and a fictitious charge $(\vq^{(1)}_m,
\vq^{(1)}_e)$ placed at the point $Q$.
On the other hand, the field to the left side of $C$ is assumed to be
given by the original particle, together with a fictitious charge
$(\vq^{(2)}_m, \vq^{(2)}_e)$ placed at the point $P$.
The boundary conditions \ethirty\ then give,
$$\eqalign{
\vqm + \vq^{(2)}_m =& - (\vq^{(1)}_m -\vqm)\cr
\vqe + \vq^{(2)}_e =& \vq^{(1)}_e +\vqe\cr
\lpto\mo L . (\vq^{(2)}_e +\vqe) - \lpoo (\vq^{(2)}_m + \vqm)
=& \lto\mo L . (-\vq^{(1)}_e +\vqe) - \loo (-\vq^{(1)}_m + \vqm)\cr
\lpto\mo L . (\vq^{(2)}_m +\vqm) + \lpoo (\vq^{(2)}_e + \vqe)
=& \lto\mo L . (\vq^{(1)}_m +\vqm) + \loo (\vq^{(1)}_e + \vqe)\cr
}
\eqn\ebcone
$$
Explicit expressions for the image charges can be found by solving these
four equations.
We are, however, interested in computing the total electric and magnetic
charges induced on the wall.
Using eqs.\ethirtyone, \ethirtytwo\ and \ebcone, these are given by,
$$\eqalign{
\Delta\vqm =& {1\over 2} (\vq^{(1)}_m + \vq^{(2)}_m) =0\cr
\Delta\vqe =& {1\over 2} (\vq^{(1)}_e + \vq^{(2)}_e)\cr
=& {(\lto)^2 - (\lpto)^2 - (\loo -\lpoo)^2\over (\lto +\lpto)^2 + (\lpoo
-\loo)^2} \vqe + {2\lto (\lpoo -\loo)\over (\lto +\lpto)^2 + (\lpoo
-\loo)^2} \mo L .\vqm\cr
}
\eqn\ethirtyfive
$$
Note that the total induced charge on the wall is independent of the
distance of the particle from the wall as long the particle is close
enough so that we can regard the wall as infinite.
As the particle approaches closer and closer to the wall, the total
induced charge gets concentrated at the point of impact.

Let us assume that after passing through the wall the particle emerges
with charge $(\vqm', \vqe')$.
A similar analysis now shows that the total charge induced on the
wall is given by,
$$\eqalign{
\Delta\vqm' =& 0\cr
\Delta \vqe'=& {(\lpto)^2 - (\lto)^2 - (\loo -\lpoo)^2\over (\lto
+\lpto)^2 + (\lpoo
-\loo)^2} \vqe' + {2\lpto (\loo -\lpoo)\over (\lto +\lpto)^2 + (\lpoo
-\loo)^2} \mo L .\vqm'\cr
}
\eqn\ethirtyeight
$$
This result can be interpreted by saying that as the particle penetrates
the domain wall, it exchanges charge with the wall.
Charge conservation then gives
$$
\eqalign{
\vqe +\Delta \vqe =& \vqe' +\Delta\vqe'\cr
\vqm +\Delta\vqm =& \vqm' +\Delta\vqm'\cr
}
\eqn\eforty
$$
Eqs.\ethirtyfive, \ethirtyeight\ and \eforty\ gives,
$$\eqalign{
\vqe' =&{\lto\over\lpto}\vqe +{1\over\lpto} (\lpoo-\loo)\mo L\vqm\cr
\vqm' =&\vqm
}
\eqn\efortyone
$$
Finally, using eq.\enineteen, we get
$$
\vqm' =\mo L\b, ~~~~ \vqe' ={1\over\lpto}(\a +\lpoo\b)
\eqn\efortythree
$$

Let us now analyse the effect of crossing the fictitious wall $D$ on the
electric and magnetic charges of the particle.
Let $\vF'_{\mu\nu}$ and $\vF''_{\mu\nu}$ be the electromagnetic fields on
the right and the left sides of this fictitious wall.
Then from eq.\efive\ we see that the boundary condition across this wall
is given by,
$$
\vF'_{\mu\nu}=(\tc\loo+\td)\vF''_{\mu\nu}+\tc\lto\mo L. \vtF''_{\mu\nu}
\eqn\efortyfour
$$
which gives,
$$\eqalign{
\vE'_{\perp, ||}=&(\tc\loo+\td)\vE''_{\perp, ||} +\tc\lto\mo
L\vB''_{\perp, ||}\cr
\vB'_{\perp, ||}=&(\tc\loo+\td)\vB''_{\perp, ||} -\tc\lto\mo
L\vE''_{\perp, ||}\cr
}
\eqn\efortyfive
$$
and the reverse relations
$$\eqalign{
\vE''_{\perp, ||}=&(-\tc\lpoo+\ta)\vE'_{\perp, ||} -\tc\lpto\mo
L\vB'_{\perp, ||}\cr
\vB''_{\perp, ||}=&(-\tc\lpoo+\ta)\vB'_{\perp, ||} +\tc\lpto\mo
L\vE'_{\perp, ||}\cr
}
\eqn\efortyfivea
$$
The total induced electric and magnetic charges on the wall are
given by $\int (\vE''_\perp - \vE'_\perp) d^2S/4\pi$ and $\int
(\vB''_\perp -\vB'_\perp) d^2S/4\pi$ respectively.
As before, we first consider a particle carrying charge $(\vqm',\vqe')$
approaching the wall from the right, and calculate the total induced
charge $(\Delta\vqm', \Delta\vqe')$ on the wall using eqs.\efortyfivea.
Then we assume that after passing through the wall, the particle carries
charges $(\vqm'',\vqe'')$ and calculate the corresponding induced charges
$(\Delta\vqm'', \Delta\vqe'')$ using eq.\efortyfive.
Finally, using the equation for charge conservation,
$$
(\vqm'+\Delta\vqm', \vqe'+\Delta\vqe')
=(\vqm''+\Delta\vqm'', \vqe''+\Delta\vqe'')
\eqn\efortynine
$$
we get
$$\eqalign{
\vqe''=&(-\tc\lpoo +\ta)\vqe' -\tc\lpto\mo L\vqm'\cr
\vqm''=&(-\tc\lpoo +\ta)\vqm' +\tc\lpto\mo L\vqe'\cr
}
\eqn\efifty
$$
Using eq.\efortythree, and that $\lpo$ is the SL(2,\Z) transform of $\lo$
by the element $\pmatrix{\ta &\tb\cr \tc &\td\cr}$ we recover
eqs.\etwentyfour, \etwentyfive.

Although this provides a rederivation of the results of the last section,
this derivation makes it clear how the charge lost by the particle is
deposited on the string.
As the particle goes farther away from the wall, the
charge spreads over wider region of the wall, thereby decreasing the
charge density induced on the wall.
When the distance of the particle from the wall is much larger than the
string size, the
total charge induced on the wall becomes negligible, showing that all the
induced charge flows back to the boundary of the wall, i.e. the string.

\chapter{Scattering of Dual Strings}

We shall now use the results derived above to study the scattering of
dual strings.
As we saw earlier, a dual string is characterized by a group element
$g=\pmatrix{a& b\cr c & d\cr}\in SL(2,\Z)$, or, equivalently, the group
element $\tg$ defined in eq.\etwelve.\foot{Note that $\tg$ remains
invariant under a change $g\to gT$.
Since it is the element $\tg$ that characterizes inequivalent strings, we
see that the elements $g$ and $gT$ describe the same string.
As we shall see, all our results will be invariant under the
transformation $g\to gT$.}
We shall first find the allowed spectrum of $\vqe$ and $\vqm$, or
equivalently of $\a$ and $\b$ defined through eq.\enineteen, for a
dual string characterized by a given element $g$.
We start with the observation that for an ordinary string
$(\a,\b)=(\vl, 0)$ where $\vl\in P$,
since these states do not carry any magnetic charge.
{}From this the allowed values of $\a$ and $\b$ for the dual string can
be found by SL(2,\Z) transformation, and are given by,
$$
\pmatrix{\a\cr -\b\cr} = \pmatrix{a & b\cr c & d\cr}\pmatrix{\vl\cr
0\cr} =\pmatrix{a\vl\cr c\vl\cr}
\eqn\efiftyfive
$$
This shows that the magnetic and electric charges of a dual string
characterized by a specific SL(2,\Z) element $g$ are related.
We shall denote the state of a dual string by the quantum numbers
$(g, \vl, \ldots)$ where $\ldots$ denote other quantum numbers which
are not of interest for our analysis.

We shall now consider two such strings, characterized by the quantum
numbers $(g_1, \vl_1)$ and $(g_2, \vl_2)$ respectively, and consider
a scattering where string 1 passes through string 2 without touching
it.
We shall assume that string 2 is a long string, where string 1 is small in
size.
The first point to note is that after string 1 passes through string 2,
it is characterized by a new group element
$$
g_1'=\tg_2^{-1}g_1
\eqn\efiftyeight
$$
\noindent Proof: Before scattering, if we go around string 1, the new
field configuration $\phi'$ is related to the original field
configuration $\phi$ through the relation $\phi'=g_1\phi$.
When string 1 crosses the fictitious wall $D$ during the process of
scattering with the string 2, we use a different coordinate system
$\psi=\tg_2^{-1}\phi$.
In this new coordinate system, the relation $\phi'=g_1\phi$ may be
expressed as,
$$
\psi'=\tg_2^{-1} \tg_1\tg_2\psi
\eqn\efiftynine
$$
Thus $\tg_1'=\tg_2^{-1}\tg_1\tg_2$.
Using the relations $\tg_1=g_1 T
g_1^{-1}$ and $\tg_1'=g_1' T g_1^{\prime -1}$ we get eq.\efiftyeight\
up to a transformation of the form $g_1'\to g_1' T$.

Using eqs.\etwentyfour, \efiftyfive\ and \efiftyeight, we see that the
electric and magnetic charge quantum numbers $\a_1'$ and $\b_1'$ of string
1 after scattering are given by,
$$
\pmatrix{\a_1'\cr -\b_1'\cr} = \tg_2^{-1}\pmatrix{\a_1\cr -\b_1}
=\tg_2^{-1} g_1 \pmatrix{\vec l_1\cr 0\cr}
=g_1'\pmatrix{\vl_1\cr 0\cr}
\eqn\esixtytwo
$$
showing that,
$$
\vl_1'=\vl_1
\eqn\esixtythree
$$

Eqs.\efiftyeight\ and \esixtythree\ determine the quantum numbers of the
string 1 after scattering.
Let us now study the quantum numbers of string 2 after scattering.
First of all, note that during this scattering process $g_2$ and $\tg_2$
remains unchanged, i.e.
$$
g_2'=g_2
\eqn\esixtyfour
$$
Conservation of electric and magnetic charge implies that the charges lost
by string 1 must be deposited on string 2.
This gives,
$$
\pmatrix{\a_2'\cr -\b_2'\cr} = \pmatrix{\a_2\cr -\b_2\cr}
+\pmatrix{\a_1\cr -\b_1\cr} -\pmatrix{\a'_1\cr -\b_1'\cr}
\eqn\esixtyfive
$$
Using eqs.\efiftyfive, \esixtytwo\ and \etwelve\ we get,
$$
\pmatrix{\a_2'\cr -\b_2'\cr} = g_2\pmatrix{\vl_2'\cr 0\cr}
\eqn\esixtyseven
$$
where,
$$
\vl_2'=\vl_2+(a_2 c_1 - a_1 c_2)\vl_1
\eqn\esixtyeight
$$
Eqs.\efiftyeight, \esixtythree, \esixtyfour\ and \esixtyeight\ describe the
final result of scattering when string 1 passes through string 2.

\chapter{Summary}

To summarize, in this paper we have studied the classical scattering of
dual strings
(strings carrying electric and magnetic charges) and have shown that there
is a definite change of quantum numbers of the string as a result of the
scattering.
The changes in the quantum numbers are determined by the quantum numbers
of the original strings, and depend on which string passed through the
other during the scattering.

The picture of magnetically charged string states that we have used is
valid for sufficiently long string states, but is not useful
for description of point like string states.
Description of such states are likely to be found in 't Hooft - Polyakov
like monopole solutions in string theory\RMONOPOLE.

\refout

\end